\documentclass[twocolumn,amsmath,amssymb,floatfix,nofootinbib]{revtex4}

\usepackage{graphicx}
\usepackage{amsmath,amssymb}
\usepackage[all]{xy}
\usepackage{epsfig}
\usepackage{graphicx}
\usepackage{epstopdf}
\newcommand{\beq}{\begin{equation}}
\newcommand{\eeq}{\end{equation}}

\newcommand{\be}{\begin{equation}}
\newcommand{\ee}{\end{equation}}
\newcommand{\kms}{ km s$^{-1}$ }

\newcommand{\apjl}{"Astrophys. J. Lett."}

\begin{document}

\title{A New Force in the Dark Sector?}
\author{Glennys R. Farrar and Rachel A. Rosen}
\affiliation{Center for Cosmology and Particle Physics,
Department of Physics\\ New York University, NY, NY 10003,USA\\
} 

\begin{abstract}
We study the kinematics of dark matter using the massive cluster of galaxies 1E0657-56.  The velocity of the ``bullet" subcluster has been measured by X-ray emission from the shock front, and the masses and separation of the main and sub-clusters have been measured by gravitational lensing.  The velocity with gravity alone is calculated in a variety of models of the initial conditions, mass distribution and accretion history;  it is much higher than expected, by at least $2.4 \sigma$.  The probability of so large a subcluster velocity in cosmological simulations estimated from the Hayashi-White fit is $\lesssim 10^{-7}$.  A long range force with strength $\approx 0.4$ - 1.2 times that of gravity would provide the needed additional acceleration.
\end{abstract}

\maketitle


Determining whether or not dark matter (DM) has long range non-gravitational interactions will help decide whether it is just a weakly interacting particle which has not yet been observed in the laboratory or a manifestation of an entirely new sector of nature.   Generically, extensions of the standard model such as supersymmetry and string theory imply the existence of scalar fields (``moduli") which are massless at all orders of perturbation theory and naturally lead to long-range interactions called ``fifth forces" for historical reasons.  To extraordinary accuracy, gravity is the only long-range force affecting baryonic matter\cite{EquivPrincCD99}, but constraints on non-gravitational forces within the dark sector are much weaker\cite{gradwohlF92}.   A fifth force may be of a Yukawa form, approximately $1/r^2$ near the source and exponentially damped beyond some range, $r_5$, which is most trivially a constant or may grow with the expansion of the Universe\cite{fpeebles03}.  In chameleon models\cite{khouryWeltmanChamNelson} the new interaction is damped by sufficiently strong concentrations of matter, so it is important to test for the existence of a fifth force on many different scales, since it may either appear or disappear at large scales.  In this work we examine the constraints on dark matter self-interactions on the Mpc scale, which can be obtained using a very special system, the colliding galaxy clusters 1E0657-56.  Thanks to the proximity and fortuitous geometry of this system, it is possible to measure the masses of the main cluster and the subcluster, and their relative velocity, and thus calculate whether the acceleration implied is consistent with the force between them being purely gravitational.  

An attractive non-gravitational force between DM concentrations is not only well-motivated theoretically, it may resolve some discomforts with conventional $\Lambda$CDM.   If the new force is long range compared to the scale of structure formation, it effectively modifies the strength of the gravitational interaction for the dark matter to be $(1 + \beta) M_1 M_2\,G_N$ and thus would accelerate structure formation.  
$\bullet$ The number of superclusters observed in SDSS data appears to be an order of magnitude larger than predicted by $\Lambda$CDM simulations\cite{einasto06}; accelerated structure formation would reduce this discrepancy.  
$\bullet$ As noted in \cite{fpeebles03}, a fifth force would tend to clear out the voids; ref. \cite{ngp04} confirms this in a simulation.  This may improve agreement with $\Lambda$CDM \cite{peeblesVoids01}, although the existence of a discrepancy is not certain\cite{mwVoids02cf2dF}.
$\bullet$ A variety of observations, for instance the lack of evidence in the Milky Way for a major merger, is hard to reconcile with the amount of accretion predicted in $\Lambda$CDM.  Accelerated structure formation reduces late-time accretion, simply because it leaves less to be accreted later\cite{ngp04}.  
$\bullet$ The number of satellites in a galaxy such as the Milky Way is predicted to be an order of magnitude larger than is observed. This ``substructure problem" is ameliorated by a 5th force, by reducing the stellar content of dwarf galaxies and making them harder to find.  This is because baryons -- not feeling the 5th force -- are relatively less-bound to dark matter concentrations than in conventional theory, reducing the amount of bound gas and lowering the star formation rate in dwarf galaxies, and increasing the tidal loss of the stars that do form.  

Structure in the Universe evolves by merger and accretion, so many large clusters of galaxies are observed to be disrupted due to a recent major merger.   A combination of factors makes 1E0657-56 unique:  \\
$\bullet$ The bullet subcluster has ``just" fallen through the main cluster for the first time, so its baryonic components -- hot gas and galaxies -- are still substantially intact. \\
$\bullet$ The subcluster and main cluster are both exceptionally massive, their relative velocity is very high, the system is nearby from a cosmological perspective ($z=0.296$), and the trajectory of the subcluster is nearly exactly in the plane of the sky.  Therefore, the supersonic shock front between the subcluster and main cluster gas is clearly visible in the X-ray images and the geometry is simple and clean enough to give an excellent determination of the relative velocity\cite{markevitch05}. 

Clowe, Markevitch and collaborators have used 1E0657-56 to exclude MOND-type alternatives to DM\cite{clowe04,clowe06}, place limits on short-range interactions of DM particles\cite{markevitch04}, and test models of X-ray production in massive clusters\cite{markevitch05}.  (See these references for a survey of earlier literature on 1E0657-56.)  With colleagues, they have also undertaken a massive campaign to improve the X-ray observations and gravitational lensing constraints.  A new 500 ks Chandra observation of the discontinuity in the density of gas across the prominent bow shock preceding the gas bullet allows precise determination of the shock Mach number and according to \cite{markevitch05} implies a shock (and bullet) velocity $4740^{+710}_{-550}\, {\rm km \, s^{-1}}$.  The errors quoted in \cite{markevitch05} were symmetrized and the central velocity rounded down(M. Markevitch, private communication); using more accurate values results in a velocity of $4740^{+710}_{-550}\, {\rm km \, s^{-1}}$ which we adopt here.  Combining weak and strong lensing gives a much improved mass distribution in the inner 500 kpc region\cite{bradac06} and leads to a separation between main cluster and subcluster mass peaks of $720\pm 25$ kpc.   A much larger weak lensing field ($34^\prime \times 34^\prime$) allows the mass distribution to be followed to larger distances.  Applying the same weak lensing analysis described in \cite{clowe04}, the new large-field weak lensing data leads to (D. Clowe, private communication) $r_{200} = 2136$ kpc, $c = 1.94$ and $r_{200} = 995$ kpc, $c = 7.12$, for the main and subclusters respectively, taking them to have spherically symmetric NFW profiles with $h_0 = 0.7, \, \Omega_M = 0.3$.   These parameters imply masses of $M_{200} = 1.5 \times 10^{15} M_{sun}$ and $M_{200} = 1.5 \times 10^{14} M_{sun}$ for the main and subclusters respectively, and virial velocity $V_{200}= 1740$ \kms for the main cluster.  A King profile fits the weak lensing data slightly better, with parameters $\rho_0 = 2.16 \times 10^6 M_{sun}\, {\rm kpc}^{-3}$, $r_c = 264$ kpc and $\rho_0 = 1.01 \times 10^7 M_{sun}\, {\rm kpc}^{-3}$, $r_c = 78$ kpc.  More general mass distributions for the main cluster were also tried; they lead to lower total mass estimates.  See \cite{CdLK04} for a discussion of systematic uncertainties in weak lensing fits.

Given the data above, we determine the most probable ``fiducial" model and use it to calculate the expected velocity of the bullet subcluster.  Variants on the fiducial model give a sense of the uncertainty in the estimate.   If the main cluster's mass distribution were static and azimuthally symmetric about the direction of motion of the subcluster, the gravitational accelerations from -720 to +720 kpc would cancel.  This trip takes less than about 0.4 Gyr, during which time mass accretion is a small effect, so our calculation is relatively insensitive to the mass distribution in the central region.  This is fortunate, because the central mass distribution is likely to be complicated due to this and earlier mergers, and may not be well described by an NFW proflle.  In the region between the central core and the virial radius, a spherical NFW density distribution fits simulations and data well on average, so we take our fiducial mass distribution for the main cluster to be a spherical NFW mass profile embedded in an otherwise homogeneous Universe of density $\rho_{M,0} (1 + z)^3$, with parameters fixed by the Clowe fit quoted above.  

The mass distribution of the main cluster evolves in time due to mergers and accretion so we need the mass distribution not only at $z = 0.296$, but at earlier times as well.  The mass accretion history (MAH) of galaxy clusters has been extensively studied\cite{MAH}.  In most cases it is well-represented by the function presented in Wechsler et al \cite{MAH} and we adopt this for the fiducial case.  To gauge the degree of dispersion we followed the actual MAHs of the 12 clusters with mass $ \geq10^{15} \, M_{sun}$, in a recent simulation of $\sim$ 80,000 galaxy clusters\cite{cwMAH05}.  In the most extreme case, the velocity using the actual MAH was 10\% higher than obtained using the mean MAH.

The cumulative effect of the multiple small gravitational deflections a test body experiences when passing through an ensemble of point masses with a distribution of velocities is known as dynamical friction.  At each position along the trajectory of the subcluster, we approximate the deceleration due to dynamical friction by the Chandrasekar formula (see, e.g., Binney and Tremaine\cite{binneyTremaine}) integrated over impact parameters in the range $r_{\rm bullet} $ to $ b_{\rm max}$ and assuming a locally Maxwellian velocity distribution. Here $r_{\rm bullet}$ is a characteristic size of the bullet which we take to be $r_{200}/c = 140$ kpc and $b_{\rm max} \equiv \left(\frac{\partial \rho}{\rho \,\partial b}\right)^{-1}$ is the scale over which the local density changes by of order one; we checked that results are only weakly sensitive to these choices.  

Note that the initial time and position are not independent because the sub-cluster is constrained to be on its first exit through the center and to have reached a radius 720 kpc at $z = 0.296$.  Hence the predicted final velocity is insensitive to the distance at which infall is presumed to start, because the acceleration is negligible until the excess mass in the main cluster above the cosmic mean background density becomes significant.  The rms peculiar velocity of galaxy clusters is $293 \pm 28$ \kms, with $< 5$ \% probability of a velocity greater than 600 \kms \cite{bahcall:clusvel96}.  Therefore we conservatively take the fiducial value of the initial infall velocity to be 300 \kms, and also consider 0 and 600 \kms. 

The fiducial model predicts a much smaller velocity than the 4740 ${\rm km \, s}^{-1}$ observed: $2950^{+ 130}_{- 90} $ ${\rm km \, s}^{-1}$, where the uncertainty range reflects the 0-600 ${\rm km \, s}^{-1}$ range of initial velocities.  To  produce the observed final velocity, an absurd initial velocity of 3135 \kms would be needed.  We varied the parameters and assumptions of the model to see how much the predicted velocity can be increased; Table \ref{results} summarizes our results.  We considered a King profile and an NFW profile truncated at the virial radius; these give still lower subcluster velocities, the larger being 2330 ${\rm km \, s}^{-1}$ for the King profile.  An N-body simulation would do a better job on dynamical friction, but as shown by the cases with no dynamical friction, improving the treatment should not substantially affect the conclusions since dynamical friction is responsible for only a 325 ${\rm km \, s}^{-1}$ decrease in the predicted velocity.  To illustrate the sensitivity to NFW parameters, we fit the surface mass density within 500 kpc from the combined weak and strong lensing analysis of Bradac et al\cite{bradac06}.  This is a promising approach but less secure for our application since it is sensitive to probable deviations from the NFW profile in the disrupted inner region, which is inessential for us.  Furthermore, it requires an extrapolation relying on a strictly NFW profile, to the larger distances relevant to our analysis.  With these parameters we obtain a larger velocity, but still much lower than observed.  The subcluster velocity is largest for a prolate NFW profile whose major axis is aligned with the direction of the subcluster.  To maximize the effect, we used $ c/a = 0.65$, corresponding to the most elongated example among the very massive halos of \cite{ShawClusHaloStat}; we re-fit the Clowe weak lensing profile to obtain 1E0657-56 parameters under this analysis.  Even this most extreme case gives a velocity 3200 ${\rm km \, s}^{-1}$, much below that observed.  Reducing the separation between bullet and center of main cluster by $1 \sigma$ to 695 kpc has a negligible effect, only increasing the predicted velocity by 35 \kms.  

\begin{table}[htdp]
\caption{Expected velocity of the bullet subcluster under various assumptions; cluster mass density model; notes.}
\begin{center}
\begin{tabular}{|c|c|c|}
\hline
$v_{\rm sub,\, f}\, {\rm ( km / s) } $   & Profile  & Notes   \\ \hline
2950  	  &  NFW, std & $c=1.94,\, r_{200} = 2.136$   	\\ \hline
2840  	   &  NFW, std & $v_{\rm in} = 0$    	\\ \hline
3080  	   &  NFW, std & $v_{\rm in} = 600$    	\\ \hline
2985  	   &  NFW, std & $r_{\rm bullet} = 695$ kpc    	\\ \hline
3275  	   &  NFW, std & no dyn fric  	\\ \hline
2330  	   &  King  & no MAH   	\\ \hline
2785  	   &  King  & no MAH, no dyn fric   	\\ \hline
3200  	   &  NFW, prolate & $c=2.16,\, r_{200} = 2.235$  	\\ \hline
3530  	   &  NFW, Bradac & $c=5.22,\, r_{200} = 2.235$   	\\ \hline
3435  	   &  prolate + unseen mass & conspiracy model  	\\ \hline
\end{tabular}
\end{center}
\label{results}
\end{table}                                                                                                                                                                                                                                                                                                                                                                                                                                                                                                                                                                                                                                                                                                                                                                                                                                                                                                                                                                                                                                                                                                                                                                                                                                                                                                                                                                                                                                                                                                                                                                                                                                                                                                                                                                                                                                                                                                                                                                                                                                                                                                                                                                                                                                                                                                                                                                                                                                                                                                                                                                                                                                                                                                                                                                                                                                                                                                                                                                                                                                                                                                                                                                                                                                                                                                                                                                                                                                                                                                                                                                                                                                                                                                                                                                                                                                                                                                                                                                                                                                                                                                                                                                                                                                                                                                                                                                                                                                                                                                                                                                                                                                                                                                                                                                                                                                                                                                                                                                                                                                                                                                                                                                                                                                                                                                                                                                                                                                                                                                                                                                                                                                                                                                                                                                                                                                                                                                                                                                                                                                                                                                                                                                                                                                                                                                                                                                                                                                                                                                                                                                                                                                                                                                                                                                                                                                                                                                                                                                                                     

Could matter not included in the NFW profile, at distances beyond the 3 Mpc of the weak lensing observation, possibly exert a sufficiently strong additional force to account for the observed velocity of the subcluster?  In the ROSAT image, 1E0657-56 seems isolated and there is no indication of other galaxy clusters in an extended filamentary structure.  Given its sensitivity limit, the ROSAT observation excludes a cluster more massive than about one-tenth that of 1E0657-56 within about 10 Mpc (A. Vikhlinin, private communication).  Other variations in the assumptions only increase the discrepancy with the prediction of gravity alone.  If the motion of the cluster is not precisely in the plane of the sky, or if it does not pass directly through the center of the main cluster, the subcluster velocity extracted from the shock front discontinuity increases because the actual distance from the center of the cluster to the leading edge of the shock is then higher so the ambient temperature and density of the pre-shock gas is lower and the density contrast is greater. (This is the reason for the asymmetric velocity errors; we thank M. Markevitch and A. Vikhlinin for discussions of this point.)  Moreover, the subcluster would be further away from the center of the main cluster, so would have been deccelerated more.  If the trajectory does not pass through the center of the main cluster, a given observed velocity implies a longer travel time and at any given radial position along the trajectory the age and mass of the main cluster are lower.  

With the above in mind, we create a ``conspiracy" model which combines those features that enhance the predicted velocity, at their 95\% confidence limits -- extreme prolate mass distribution aligned in the direction of motion, infall velocity of 600 \kms, and a static mass $1.5 \, \times 10^{14} M_{sun}$ located 3 Mpc away from the center of the main cluster in the direction of motion of the subcluster -- and we reduce the observed separation by $1 \sigma$ to 695 kpc.   This conspiracy model predicts an observed velocity of 3435 \kms.

If the anomalously large observed velocity of the bullet subcluster is the result of a new Yukawa-like interaction in the dark sector whose range is larger than several Mpc, its effect on the bullet dynamics is to multiply the gravitational acceleration by the factor $(1+\beta)$.  For the fiducial case, the value of $\beta$ which would bring the predicted velocity into agreement with the  central value of the observed velocity for $r_5 = \infty$ is $\beta = 1.2$, and for the conspiracy case is $\beta = 0.8$; reducing the final velocity by $1 \sigma$ to 4190 \kms, $\beta = 0.4$ is needed in the conspiracy model.  These values are generally consistent with the limits on a fifth force found in \cite{gradwohlF92} which were only rough estimates due to the much more primitive state of observational astrophysics at that time.  Just as the dynamics of 1E0657-56 is insensitive to the mass distribution inside $\sim$720 kpc, it is likewise insensitive to a fifth force in this region and would be incapable of discriminating between a simple Yukawa interaction and a chameleon interaction which is suppressed in the center of the cluster.


Hayashi and White in \cite{hwRareBullet06} (HW) have recently taken a complementary approach to the direct dynamical analysis presented above, investigating how frequently subclusters in the Millenium Run simulation\cite{springelMR} have comparable characteristics to the bullet.  Using the bullet data prior to 2005, HW find the likelihood of a bullet-like subcluster to be about 1 in 500.  The new observations reduce this because $v_{\rm bullet}/V_{200}$ increases from 1.9 adopted by HW to $ 4740/1740 = 2.7 $.  HW fit the distribution of velocities of most massive sub-clusters divided by the virial velocity of their host, $V_{\rm sub}/V_{200} $, to obtain:  ${\rm log} \frac{N_1(> V_{\rm sub})}{N_{\rm hosts} } = - \left(\frac{V_{\rm sub}/V_{200} }{ v_{10\%} } \right)^\alpha$, with $v_{10\%} = 1.55$ and $\alpha = 3.3$ at $z = 0.28$.  Thus increasing the velocity ratio from 1.9 to 2.7 reduces the likelihood by a factor $ 4\times 10^{-5}$ so the HW likelihood estimate becomes $0.8 \times 10^{-7}$.   This remains small even when measurement errors are included: reducing $v_{\rm bullet}$ by $1 \sigma$ to $4190$ \kms and increasing $M_{200}$ by 20\% (a very conservative estimate of the uncertainty, obtained by augmenting the quoted 16\% error on the first small field, low statistics weak lensing mass determination\cite{clowe04} by 25\% for systematic errors) gives $v_{\rm bullet}/V_{200} = 2.2$ for a $1.4 \times 10^{-4}$ likelihood.  These are only estimates because the observed ratio $v_{\rm bullet}/V_{200} $ is higher than any observed in the simulation so the likelihood values rely on the HW extrapolation formula.  Nonetheless, they make clear that the bullet is rare enough it should arouse suspicion and deserves further study.

Other manifestations of a non-gravitational force between dark matter concentrations are often masked by our lack of independent information on the underlying DM system.  An important and clean test would be to stack SDSS clusters to obtain their total mass distribution from weak lensing, and to map the peculiar velocity distribution for the same stack of clusters.  If there is a fifth force, the maximum excursion in the velocity distribution will be larger than expected \cite{CAIRNS03}, although a simulation would be needed to interpret the results given complications such as dynamical friction, stripping, etc.  On a sub-galactic scale, Kesden and Kamionkowski have pointed out that a difference in acceleration between DM and stars would change the distribution of stars in the tidal tails of the Sagittarius dwarf galaxy\cite{kesdenKamPRL}.  They argue that the method should be sensitive to $\beta \gtrsim 0.1$, but until simulations with gravity alone can consistently describe all observations it is premature to claim any limits on $\beta$.  In particular, it is intriguing that the difficulty of reconciling the observed line-of-sight velocities of the stars in the leading stream, which implies a prolate Dark Matter halo, with the precession of the debris orbital plane, which implies an oblate halo\cite{ljmApJ05}, might be resolved by an additional attractive force.


We have shown that it is very difficult to reconcile the reported velocity of the bullet subcluster with the observed matter distribution which accelerated it.  With a downward error of 550 \kms, the subcluster velocity is nominally $3.25 \sigma$ higher than the fiducial prediction of $2950$ \kms.  A conspiracy model, requiring 95\% CL extremes of initial infall velocity, elongation ratio of a prolate mass distribution, a hypothetical additional mass concentration at the estimated maximum level allowed by weak lensing and ROSAT, and a $1 \sigma$ smaller separation, predicts a velocity of 3435 \kms, still falling short of observation by $2.4 \sigma$.   Clearly, further effort to reduce uncertainties in all relevant quantities is warranted.  Reducing reliance on simple parameterizations of the mass distribution would be particularly helpful.  Adding more red-shifts for the arcs and extending the strong+weak lensing method to larger radii, as is underway, will strengthen the lensing determination of the mass distribution.  Complementary methods of estimating the mass distribution, such as the X-ray temperature-intensity method of ref. \cite{X-raymass} should be pursued.  A holistic treatment using simulations to integrate the lensing information on the dark matter distribution with information on the gas and stellar dynamics from X-ray and optical data, could substantially improve our understanding of this important system and is underway \cite{sf07}.  

If the discrepancy reported here between predicted and observed dynamics of the bullet subcluster is substantiated by refined observations and analysis, and confirmed in other systems, it would imply the existence of a long-range, non-gravitational force within the dark sector.  This would have profound implications for particle physics and
would provide a unique window onto an entirely new side of particle physics involving, perhaps, extra dimensions or the most basic degrees of freedom of string theory. The next step would be to determine the range of the force, how it varies with the scale factor of the Universe, and whether the Yukawa description is applicable or breaks down in dense concentrations of dark matter -- characteristics of a more subtle dynamics than simple exchange of an ultralight scalar particle, which are in principle accessible through careful systematic study of the dark sector dynamics.  A consequence of a long-range non-gravitational force in the dark sector would be that dark matter will be difficult or impossible to detect in laboratory experiments, 
because otherwise quantum corrections would give rise to an unobserved equivalence-principle-violating interaction of ordinary matter; details will be reported elsewhere.

{\bf Acknowledgments:}  We are grateful to D. Clowe, M. Markevitch and M. White for sharing unpublished results, and to them and M. Bradac, J. Cohn, D. Croton, G. Gabadadze M. Gladders, E. Hayashi, H. Hoekstra, K. Johnston, A. Kravtsov, E. Sheldon, V. Springel,  A. Vikhlinin, and S. White for valuable discussions.  This research has been supported in part by NSF-PHY-0401232.


\end{document}